\documentclass[prd,twocolumn,showpacs,superscriptaddress]{revtex4}
\usepackage{amsmath, amssymb, graphics, epsfig}
\usepackage[dvipsnames]{xcolor}

\newcommand{\mathsym}[1]{{}}

\newcommand{\AP}{Alcock-Paczy\'{n}ski }
\usepackage{xspace}

\newcommand{\lcdm}{$\Lambda$CDM\xspace}
\newcommand{\vide}{\tt VIDE\normalfont\xspace}
\newcommand{\zobov}{\tt ZOBOV\normalfont\xspace}

\begin{document}

\title{Counting voids to probe dark energy}



\author{Alice Pisani}
\email{pisani@cppm.in2p3.fr}
\affiliation{Centre de Physique des Particules de Marseille, Aix-Marseille Universit\'e, CNRS/IN2P3, Marseille, France}
\affiliation{Sorbonne Universit\'es, UPMC (Paris 06), UMR7095, Institut d'Astrophysique de Paris, 98bis Bd. Arago, F-75014, Paris, France}
\affiliation{CNRS, UMR7095, Institut d'Astrophysique de Paris, 98bis Bd. Arago, F-75014, Paris, France}

\author{P. M. Sutter}
\affiliation{INFN - National Institute for Nuclear Physics, via Valerio 2, I-34127 Trieste, Italy}
\affiliation{INAF - Osservatorio Astronomico di Trieste, via Tiepolo 11, 1-34143, Trieste, Italy}
\affiliation{Center for Cosmology and AstroParticle Physics (CCAPP), The Ohio State University, 191 West Woodruff Avenue, Columbus, Ohio 43210, USA}

\author{Nico Hamaus}
\affiliation{Sorbonne Universit\'es, UPMC (Paris 06), UMR7095, Institut d'Astrophysique de Paris, 98bis Bd. Arago, F-75014, Paris, France}
\affiliation{CNRS, UMR7095, Institut d'Astrophysique de Paris, 98bis Bd. Arago, F-75014, Paris, France}

\author{Esfandiar Alizadeh}

\affiliation{Dept. of Physics, Univ. of Illinois at Urbana-Champaign, 1110 West Green Street, Urbana, IL 61801}

\author{Rahul Biswas}
\affiliation{Department of Astronomy and eScience Institute, University of Washington, Seattle}

\author{Benjamin D. Wandelt}
\affiliation{Sorbonne Universit\'es, UPMC (Paris 06), UMR7095, Institut d'Astrophysique de Paris, 98bis Bd. Arago, F-75014, Paris, France}
\affiliation{CNRS, UMR7095, Institut d'Astrophysique de Paris, 98bis Bd. Arago, F-75014, Paris, France}
\affiliation{Depts. of Physics and Astronomy, Univ. of Illinois at Urbana-Champaign, Urbana, IL 61801, USA}
\author{Christopher M. Hirata}
\affiliation{Center for Cosmology and AstroParticle Physics (CCAPP), The Ohio State University, 191 West Woodruff Avenue, Columbus, Ohio 43210, USA}

\begin{abstract}
We show that the number of observed voids in galaxy redshift surveys is a sensitive function of the equation of state of dark energy. Using the Fisher matrix formalism we find the error ellipses in the $w_0-w_a$ plane when the equation of state of dark energy is assumed to be of the form $w_{CPL}(z)=w_0 +w_a z/(1+z)$. We forecast the number of voids to be observed with the ESA Euclid satellite and the NASA WFIRST mission, taking into account  updated details of the surveys to reach accurate estimates of their power. The theoretical model for the forecast of the number of voids is based on matches between abundances in simulations and the analytical prediction. To take into account the uncertainties within the model, we marginalize over its free parameters when calculating the Fisher matrices. The addition of the void abundance constraints to the data from Planck, HST and supernova survey data noticeably tighten the $w_0-w_a$ parameter space. We thus quantify the improvement in the constraints due to the use of voids and demonstrate that the void abundance is a sensitive new probe for the dark energy equation of state.
\end{abstract}

\maketitle

\section{Introduction}

The measurement of luminosity distances from  supernovae of type Ia \citep{Riess1998,Perlmutter1999} gives observational evidence for the accelerated expansion of the Universe. To account for such acceleration, the standard cosmological model relies on the existence of a mysterious component: dark energy.
Different models of dark energy are viable, such as models where the Universe is permeated by a constant energy density (e.g. a cosmological constant with $w_0=-1$) or models of dynamical dark energy. A precise measurement of the equation of state of dark energy allows to distinguish among such models. 

The large scale structure of the Universe depends on the cosmological model, thus its study allows to trace the rate of expansion of our Universe. For instance, the number of galaxy clusters depends on the properties of dark energy: a comparison of observational counts with analytical predictions permits to constrain the dark energy equation of state \citep{Basilakos2010}. While this measure relies on the high density regions of the universe, the use of the low density ones, such as cosmic voids, is also sensitive to the properties of dark energy. 
Additionally, while cluster abundances can be problematic because of the high intrinsic scatter in their mass-observable relationships, voids promise to suffer from this uncertainty in a milder (and different) way, as their volumes can be observed and compared to theory directly.

Cosmic voids have remained fairly unexplored until recent times, due to the difficulty in acquiring sufficient data from sparsely populated regions of the Universe. Nowadays, as modern surveys such as the SDSS map out the cosmic web in great detail, entire void catalogs are being compiled \cite{Pan2012,Sutter2012a,Nadathur2014} and used in a wide range of cosmological applications, spanning from the \AP test \citep{Sutter2014,Hamaus2014b} to the measurement of the integrated Sachs-Wolfe effect \citep{Ade2014}. Analogously to galaxy cluster number counts, here we propose to use void number counts to constrain the cosmological model: the number of observed voids is sensitive to the dark energy equation of state. This means that simply counting voids in galaxy surveys enables us to constrain the properties of the latter. 

Theoretical models based on the Press-Schechter formalism \cite{Press1974} adapted for the case of voids (known as the Sheth and Van de Weygaert model \cite{Sheth2004}) and the recent observational work with voids allows in principle to forecast the number of voids that should be observed by a particular survey and with a given cosmological model. While recent works have extended the original Sheth and Van de Weygaert model (such as \cite{Paranjape2012, Jennings2013, Achitouv2013}), for this work we will consider the original model, in the framework of which we are qualitatively able to match simulations mimicking the tracer number density of upcoming surveys (as explained throughout the paper). 
Thus, the predicted abundance of observed voids for each galaxy redshift survey gives the power of the survey to constrain the dark energy equation of state through void counting. To reach this goal, we use the Fisher matrix formalism \cite{Fisher1935,Verde2010}(as it has been done for different probes by \cite{Biswas2010} and \cite{Lavaux2012}).

In this paper we focus on two upcoming surveys, Euclid \cite{Laureijs2011} and WFIRST (Wide Field Infrared Space Telescope) \cite{Spergel2013,Spergel2013a}, to establish their expected power to constrain dark energy in terms of void abundance.
The two surveys are in some sense complementary. The Euclid satellite is a Medium Class mission of the ESA-led Cosmic Vision 2015-2025 that will cover $15000~\rm deg^{2}$ of the sky. It will give pseudo-spectroscopic redshifts of tens of millions of galaxies up to high redshift. On the other hand, WFIRST --- a NASA-led mission --- will cover a smaller portion of the sky, $2000~\rm deg^{2}$, but with a higher sampling density of galaxies. The two surveys thus present two complementary strategies, on the one side a wide but shallower survey (Euclid); on the other side a deeper but narrower field (WFIRST). 

Naturally, one might ask which approach can provide the largest number of voids. Because cosmic voids exhibit a hierarchical structure \cite{Sheth2004,Aragon-Calvo2013}, an increase in the density of galaxies (a higher resolution sampling of the large-scale structure of the Universe) allows resolving voids of smaller size at each redshift. On the other hand, a wider field also increases the number of observed voids, so it is not obvious which survey strategy is best suited for maximizing the number of observed voids. 

In this paper we aim to answer this question by presenting a forecast for the number of voids to be detected by Euclid and WFIRST. We also establish the constraining power from their void abundances in view of the dark energy equation of state. To this end, we consider the Chevallier-Polarski-Linder \citep{Chevallier2001,Linder2003} parametrization for the dark energy equation of state:
\begin{equation}
w_{CPL}(z)=w_{0}+w_{a}\frac{z}{z+1}
\end{equation}

Once the number of observed voids is known, the Fisher matrix formalism can provide error ellipses in the $w_{0}-w_{a}$ plane. The Fisher formalism gives an estimate of the constraints to be expected from cosmological probes in terms of area in parameter space.

Recent work on observational data (e.g. the SDSS DR7 and DR9 data, see \cite{Sutter2012a,Sutter2014e}) sheds more light on the behaviour of voids that are defined in the distribution of galaxies (as opposed to dark matter \cite{Sutter2014c, Leclercq2014}), thereby allowing a more robust estimation of void abundances based on the use of simulations and mock galaxy catalogs that are calibrated to reproduce the properties of the data in the most realistic way possible. 

A number of cosmological tests based on void statistics 
have been proposed, e.g. \AP test on void stacks \cite{Lavaux2012,Sutter2012b,Sutter2014,Hamaus2014b}, tests of coupled dark energy-dark matter models \cite{Sutter2014b}, tests of modified gravity models \cite{Spolyar2013,Cai2014a,Zivick2014}, constraints from CMB lensing with cosmic voids \cite{Chantavat2014}.
To understand the potential of void-based cosmology with Euclid and WFIRST requires realistic estimates of void abundances based on updated details on both missions, such as those we present in this paper.

The paper is organized as follows: in Section 2 we present the theoretical calculation of the void abundance. In Section 3 we discuss the determination of the minimum void radius to be considered useful for our analysis. Section 4 presents the Fisher matrix analysis and in Section 5 we describe the simulations used for this work. Finally in Section 6 we present our results and conclude in Section 7.

\section{Void Number Calculation}

In order to estimate the constraining power from voids that can be observed with future surveys, we need to have a theoretical prediction of their expected abundance. This, in turn, requires a theoretical model for the formation of voids. One possibility to define the formation of a void is to consider the spherical expansion of a top-hat perturbation, analogously to the case of halo formation by spherical collapse \cite{Blumenthal1992}, but with opposite sign. The moment of void formation can be defined as the evolutionary stage where two adjacent radial shells start crossing each other, leading to the formation of a ridge \cite{Sheth2004}. This consideration allowed Sheth and Van de Weygaert \cite{Sheth2004} to construct a model for the distribution of void sizes. We take this model as a starting point for determining void abundances.

Following the Press-Schechter formalism \cite{Press1974}, we calculate the abundance of voids using
\begin{equation}
\frac{M^2 \, n(M,z)}{\bar{\rho}}\frac{\mathrm{d}M}{M} = \nu f(\nu) \frac{\mathrm{d}\nu}{\nu}\;,
\label{equ:PS}
\end{equation}
where  $M$  is the void mass, $\bar{\rho}$ the background density, and $n(M,z)$ the number density of voids of given mass and redshift. According to Sheth and Van de Weygaert \cite{Sheth2004}, assuming Gaussian initial conditions, the fraction $f(\nu)$ of mass that has evolved into voids can be approximated as: 
\begin{equation}
\nu f(\nu ) \approx \sqrt{\frac{\nu}{2 \pi}} \exp (-\nu /2)\;, \label{eq: nu}
\end{equation}
where $\nu =\frac{\delta _v^2}{\sigma ^2(M)}$, $\delta_v$ is the critical underdensity of void formation, linearly extrapolated to the time when shell-crossing takes place at the void edge, and $\sigma^2(M)$ is the variance of linear density fluctuations filtered on a scale $R=(\frac{3M}{4\pi\bar{\rho}})^{-1/3}$. We anticipate here that we consider equation \ref{eq: nu} in the case were the two-barrier distribution reduces to a single barrier, a particular case described in \cite{Sheth2004} where the void-in-cloud process is neglected, as detailed in the next Section. 

Recent work measured the abundance of voids in galaxy surveys (\textit{e.g.~} \cite{Sutter2014e}) and compared it to mock catalogs from simulations with properties matching the data. The Sheth and van de Weygaert excursion set model can be used to model the void abundances accurately, as long as the value of $\delta _v$ is taken as a free parameter. While this is a phenomenological approach, it uses the power of the theoretical description combined with the added realism by fitting a parameter to simulations matching observations. Indeed the distribution of voids needs to be validated with simulations in order to be reliable \cite{Jennings2013}. 

Thus, in the redshift range of interest to observe voids, we have used sub-sampled simulations matched to the tracer density of Euclid and WFIRST to tune this parameter optimally. The value of $\delta _v$ obtained with this methodology is $-0.45$, we find this value to be stable when varying the cosmological parameters. The value of $\delta _v$ differs from the Sheth and van de Weygaert prediction because watershed voids are not spherical and do not exhibit uniform shell-crossing. As mentioned, recent works show that some improvements can be obtained~\cite{Jennings2013} by validating the void distribution with mocks. The use of simulations tuned to observed data \cite{Sutter2014e} thus enhances the robustness of the abundance estimation we perform.

Additionally, since the value of $\delta _v$ might partially depend on the void definition, we use the same void finder (\vide \cite{Sutter2014a}, based on \zobov \cite{Neyrinck2008}) for the treatment of existing data and simulations with similar properties --- to minimize bias due to the void definition. This is only a consistency choice, since generally, we do not expect void abundances to change much for void finders based on a similar concept as \vide (namely a tessellation of the tracer distribution followed by a watershed transform to detect the structure of the cosmic web).

In Equation \ref{eq: nu} we used $\sigma$, the linearly extrapolated variance of density fluctuations $\delta$ smoothed on the filtering scale $R_{\mathrm{Lag}}$ by the Fourier transform of the window function for a top-hat filter $\tilde{W}(x)=3/x^3 \left[ \sin(x)-x \cos(x) \right]$. Considering the matter power spectrum $P_{\delta}(k,a)$, we write the definition of $\sigma$: 
\begin{equation}
\sigma^2(M,a) = \int_0^{\infty} \frac{k^3 P_{\delta}(k,a)}{2 \pi^2} \left| \tilde{W}(k R_{\text{Lag}}(M)) \right|^2 \frac{\mathrm{d}k}{k}
\label{eq: sigma}
\end{equation}

The calculation of $\sigma$ depends on cosmology, thus it can be used to constrain cosmological parameters through the observed abundance of voids in a survey. 
We can change variables from the mass of the void to its radius (either Lagrangian or Eulerian) by using mass conservation inside the void:

\begin{equation}
M = \frac{4\pi}{3} R_{\text{Lag}}^3 \, \bar{\rho} = \left(1+\Delta _V\right) \frac{4\pi}{3} R_{\text{Eul}}^3 \, \bar{\rho} ,
\label{equ:EandL}
\end{equation}
where $\Delta_V$ is the underdensity of the matter inside the void. From the spherical collapse model one finds $\Delta_V\simeq-0.8$ at the time of shell crossing. Since the expansion of the void reduces considerably after shell-crossing
(see \cite{Sheth2004, Furlanetto2006}) we will assume this value of $\Delta_V$ for all voids irrespective of when the shell formed around them. It should be noted that we have written the Eulerian radius in the comoving form, hence the background density is constant and equal to its present day value. To prevent confusion we will keep the Eulerian and Lagrangian subscripts for the radii throughout the paper.
The observable quantity we measure in a galaxy redshift survey is the number of voids larger than a given radius $R^{\mathrm{min}}_{\mathrm{Eul}}$ located in a specified redshift interval. It can be found as

\begin{equation}
N_{e} = \int_{z}^{z+\Delta z} \mathrm{d}z \int_{R^{\mathrm{min}}_{\mathrm{Eul}}}^{\infty} \mathrm{d}R_{\text{Eul}} \int_{\Omega_{\text{survey}}} \mathrm{d} \Omega \, n(R_{\text{Eul}},z) \frac{\mathrm{d}V}{\mathrm{d}z \mathrm{d}\Omega}\;, 
\label{equ:Ne}
\end{equation}
where $\mathrm{d}V$ is the comoving volume element and the angular integration is over the angular size of the survey. The differential volume element is a function of cosmology through the Hubble rate. 

Considering Equations \ref{eq: sigma} and  \ref{equ:Ne}, the number of voids will depend on the cosmological model through $\sigma$ and the differential volume element. Additionally, the minimum radius of voids considered is a relevant quantity that will affect the forecast of void abundance. Its choice will be discussed in the next Section.

\section{Minimum Void Radius Determination}

The constraining power of a void survey crucially depends on the minimum observable void radius. Particular care must be taken to determine this minimum radius for the voids in the survey: the abundance of voids increases rapidly with decreasing radius (analogously to the abundance of clusters). We consider two different criteria for defining the minimum void radius and take into account the most stringent \cite{Biswas2010}, \textit{i.e.~} the largest radius of the two.

The first and simplest criterion is based on the mean particle separation of galaxies in the survey. Taking into account the features of the survey, we consider as a minimum radius of a void twice the mean particle separation. Empty spaces among galaxies might not correspond to true voids in the dark matter distribution, as they may simply result from the discrete sampling of the underlying distribution (shot noise). 

Voids can be defined reliably down to the mean particle separation, but considering twice the mean particle separation $2R_{mps}$ also guarantees a small impact of peculiar velocities in redshift space \cite{Pisani2014b}. Indeed, voids below that limit can be washed out or created by the effect of peculiar velocities. Therefore, to increase our confidence that an observed underdensity in the galaxy distribution is a true void and to limit systematic effects due to velocities, we demand that the radius of the void must be larger than twice the mean particle separation $R^{\text{Shot}}_{\text{Eul}}=2.0\times (1/\bar{n}_{\text{gal}})^{1/3}=2.0\times R_{mps}$. 

The comoving number density of galaxies $\bar{n}_{\text{gal}}$ depends on survey specifications, such as the maximum apparent luminosity observable or the galaxy luminosity function (see \citep{Biswas2010} for more details). While the Euclid redshift survey has undergone some changes recently (which might lead to updated values); to refer as much as possible to realistic values for the galaxy number densities of the survey, for the present paper we will use as representative for each redshift bin the values reported by the \textit{Review of the Euclid Theory Working Group} \cite{Amendola2013}. For the WFIRST satellite, while our results do not change much using the values in the \textit{Wide-field infrared survey telescope-astrophysics focused telescope assets WFIRST-AFTA final report} \cite{Spergel2013a}, we use updated values from the most recent \textit{Wide-Field InfrarRed Survey Telescope-Astrophysics Focused Telescope Assets WFIRST-AFTA 2015 Report} \cite{Spergel2015}.

A second threshold for the minimum observable void size is set by the \textit{void-in-cloud} effect \cite{Sheth2004}. It is due to the fact that small voids tend to be located inside surrounding overdensities, and therefore will disappear with the gravitational collapse of the overdense region around them. In such cases the Press and Schechter formalism must be extended. As proposed by Sheth and Van de Weygaert \cite{Sheth2004}, a two-barrier excursion set model should be used --- the need for two parameters for the void size distribution is unique to the void case. It involves solving a diffusion equation with two absorbing barriers instead of one, one for halos and one for voids.

The void-in-cloud effect leads to a turn-over radius below which the number density of voids decreases towards smaller radii. For our analysis, as discussed in Section 2, we want to avoid this regime, so we consider voids with radii larger than the Lagrangian void-in-cloud radius for which we have $\sigma(R^{\text{VinC}}_{\text{Lag}},z)\simeq 1$. Such voids are unlikely to be embedded in an overdense region, hence giving a framework where the Press and Schechter formalism can be applied (only the smallest voids are subject to collapse \cite{Sheth2004},\cite{Sutter2014d}). The analysis of velocity profiles confirms this scenario \cite{Weygaert1993,Hamaus2014,Paz2013}. We thus calculate this radius, convert it from Lagrangian to Eulerian radius with equation \ref{equ:EandL} and use it as the second criterion (it should be noted that this criterion is rarely relevant, since it is usually less constraining than the criterion of twice the mean particle separation).

At each redshift bin, the respectively higher value among $2R_{\mathrm{mps}}$ and $R^{\text{VinC}}_{\mathrm{Eul}, \sigma \simeq1}$, thus defined as:

\begin{equation}
R_{\text{Eul}}^{\text{min}} = \max (R_{\text{Eul}}^{\text{VinC}},2R_{\mathrm{mps}})
\end{equation} 
is used as the minimum radius to obtain the abundance of voids. Thus, at each redshift we consider the most stringent of the two constraints, one taking into account the survey's features ($2R_{\mathrm{mps}}$), the other depending on cosmology ($R^{\text{VinC}}_{\mathrm{Eul}, \sigma \simeq1}$). 

In the next Section we present the Fisher formalism used in this analysis.

\section{Fisher analysis}

The Fisher formalism is a method to determine a lower bound on the expected uncertainties with which cosmological parameters can be determined. We rely on this formalism to estimate the power of surveys in constraining cosmological models with void abundances.

We discretize the survey into separate redshift bins and assume that the number of observed voids in any bin is a random draw from a Poisson probability distribution. That is, if the expected number of voids in the bin from the theory is $N_e$, whose calculation was the subject of Section II, then the observed number of voids, $N_o$, has the probability distribution

\begin{equation}
p(N_o|N_e)=\frac{N_e^{N_o}e^{-N_e}}{N_o!} .
\end{equation}

We assume that the drawings from two different redshift bins are uncorrelated so that the probability to find $N_{o,1}$ in the first bin, $N_{o,2}$ in the second bin and so on, simply is:

\begin{equation}
L(N_{o,1},N_{o,2}, \text{...}|N_{e,1},N_{e,2}, \text{...}) = \underset{a=1}{\overset{N_{\text{bin}}}{\prod}} \frac{N_{e,a}{}^{N_{o,a}}e^{-N_{e,a}}}{N_{o,a}!}.
\label{equ:likelihood}
\end{equation}

\noindent The Fisher information matrix is defined as:

\begin{equation}
F_{mn} = E \left[\frac{\partial \ln L}{\partial \theta _m} \frac{\partial \ln L}{\partial \theta _n} \right]
\end{equation}

\noindent where $E$ stands for the expectation value and the $\theta _i$ denote the cosmological parameters. For the case of the Poisson distribution we will have:

\begin{equation}
\frac{\partial \ln L}{\partial \theta _m} = \underset{a}{\sum} \left(\frac{N_{o,a}}{N_{e,a}}-1 \right) \frac{\partial N_{e,a}}{\partial \theta _m}
\end{equation}

Considering Equation \ref{equ:likelihood}, using $E\left[N_{o,a}\right]~=~N_{e,a}$ and the fact that for the Poisson distribution

\begin{eqnarray}
\nonumber E\left[N_{o,a}N_{o,b}\right] & = & E\left[N_{o,a}\right] \delta_{ab} + E\left[N_{o,a}\right]E\left[N_{o,b}\right] \\
& = & N_{e,a} \delta_{ab} + N_{e,a}N_{e,b} ,
\end{eqnarray}

\noindent it immediately follows \citep{Holder2001} that:

\begin{equation}
F_{mn} = \underset{a}{\sum} \frac{\partial N_{e,a}}{\partial \theta _m}\frac{\partial N_{e,a}}{\partial \theta _n}\frac{1}{N_{e,a}} \label{eq: Fisher}
\end{equation}

\noindent where the right hand side is calculated for a fiducial set of values of the parameters $\theta$. We note that, when deviating from fiducial cosmology, the inferred void radius will change. We consider the modification of the void radius when the adopted cosmology is not the fiducial one as:

\begin{equation}
R_{\text{Eul}} = R_{\text{Eul, true}}\Bigg( \frac{\mathrm{d}V_{\text{fid}}/(\mathrm{d}z \mathrm{d}\Omega)}{\mathrm{d}V_{\text{true}}/(\mathrm{d}z \mathrm{d}\Omega)} \Bigg) ^{1/3} \label{eq: inferred radius}
\end{equation} 

The variation of the radius of voids if a cosmology different than the fiducial is assumed needs to be taken into account when performing the Fisher analysis, since it affects the derivatives of equation \ref{eq: Fisher}. We note that $V_{\text{true}}$ and $V_{\text{fid}}$ need to be in the same units. While correctly taking into account this effect (considering equation \ref{eq: inferred radius} when we calculate the numerical derivatives for $N_{e}$ in equation \ref{eq: Fisher}), for the present analysis we are neglecting further effects (such as the \AP effect), which should be quantified in future analysis with mocks based on Halo Occupation Distribution models (that promise to be more accurate when further details about Euclid and WFIRST will be available) and for the application of this method to data.

The Cramer-Rao theorem states that any unbiased estimator of the parameters $\theta_A$ will have a covariance matrix larger than, or at best equal to, the inverse of the Fisher matrix.

Thus the Fisher matrix formalism can give an estimate of the constraints to be obtained from cosmological probes in terms of area in the parameter space. 

While more sophisticated techniques exist to estimate such constraints \cite{Wolz2012}, the Fisher matrix analysis is adapted for a first application of our simulation-calibrated number functions method. 

\begin{figure}
\centering
\epsfig{file=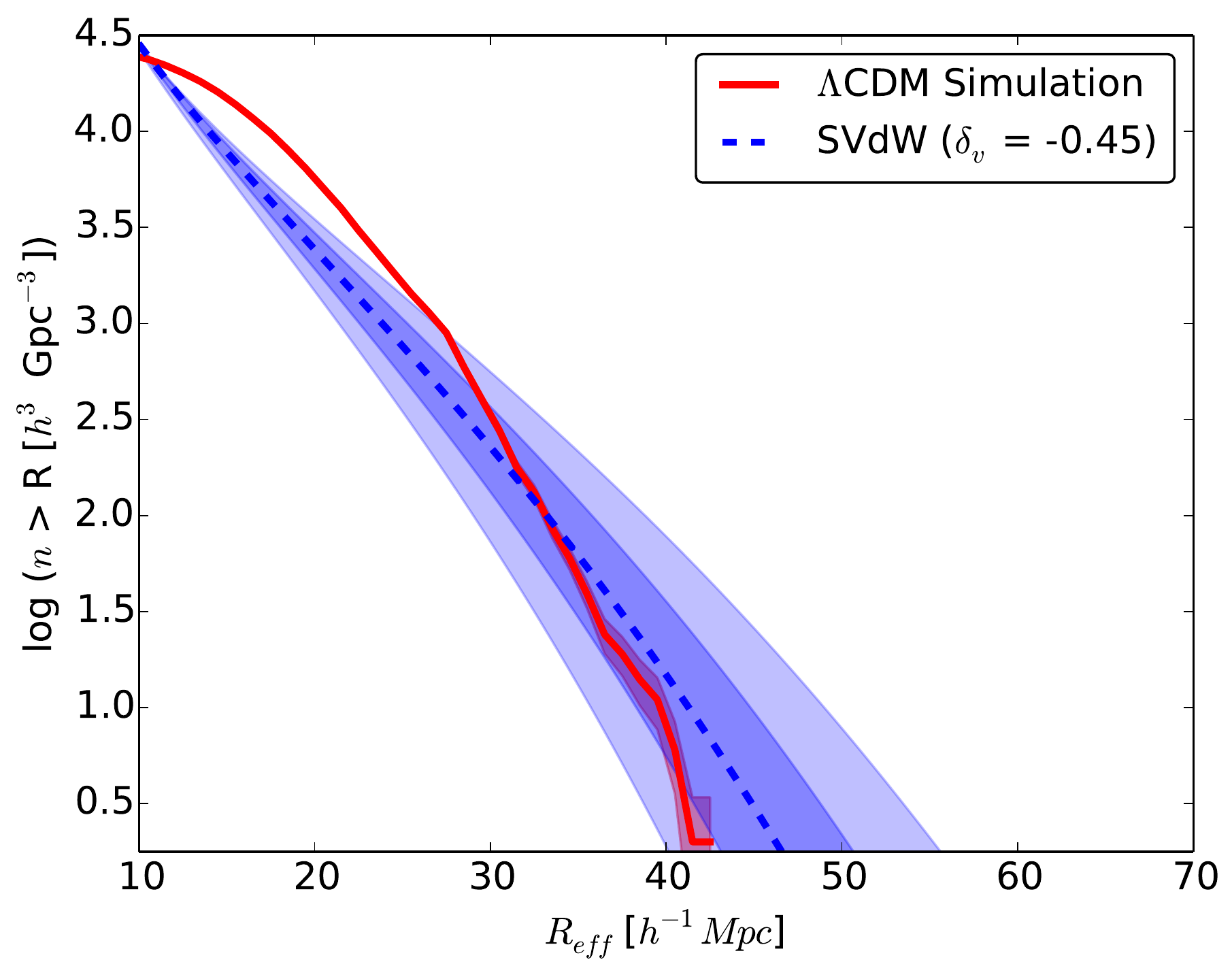,width=1\linewidth,clip=, angle=0}  
\caption{We represent the prediction from the Sheth and Van de Weygaert model matched to the simulation. The shaded areas show the effects of the variation of the value of $\delta_{v}$ (varied by 0.05 and 0.1 in the Figure) for the \lcdm case. The marginalization over $\delta_{v}$ allows to take into account the fact that the Sheth and Van de Weygaert fit is qualitative, and only roughly matches simulations due to the discreteness of the galaxy distribution as opposed to a smooth density field. \label{fig: Abundances0}}
\end{figure}

\begin{figure} 
\centering
\epsfig{file=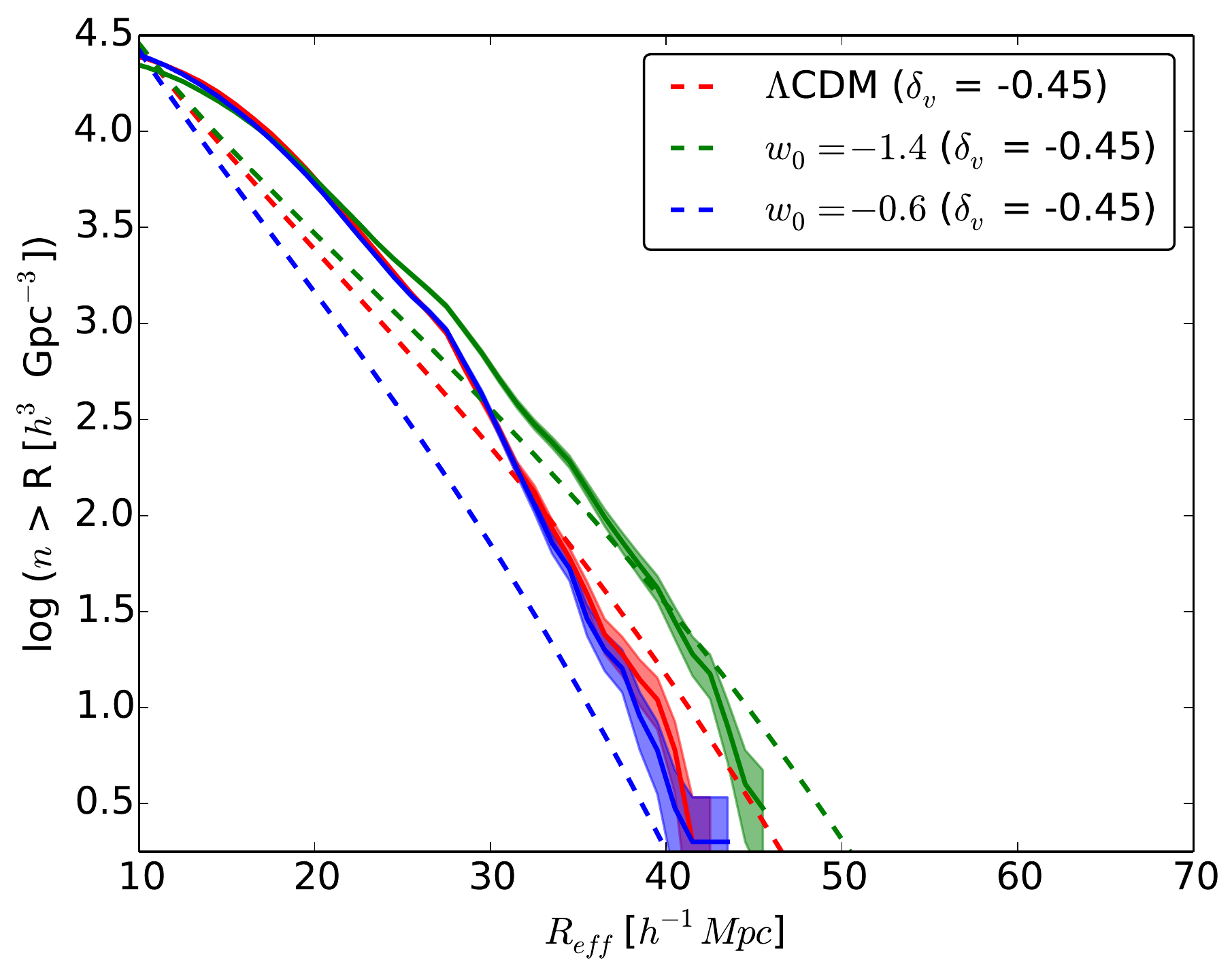,width=1\linewidth,clip=, angle=0}  
\caption{Matching of the abundance of voids in simulations for models with different $w_0$. The lines show the abundances from simulations for models with different $w_0$ (and fixed $w_a$). The prediction from the Sheth and Van de Weygaert model are represented with dashed lines.
\label{fig: Abundances1}}
\end{figure}

\section{Simulations}

As discussed in Section I, we used simulations to tune the parameter $\delta_{v}$. We use simulations of sub-sampled dark matter and use the void finding technique described in \cite{Sutter2014a}. While we leave for future work the tuning of $\delta_{v}$ with mocks constructed using Halo Occupation Distribution models (which promise to be more accurate when further details about Euclid and WFIRST will be available), the subsampling of simulations allows us to mimic the number density of tracers in the two surveys. 

For the simulations, we have used an adaptive treecode N-body method whose operation count scales as N log N in the number of particles, the 2HOT code \cite{Warren2013}. With the use of a compensating smoothing kernel for small-scale force softening \cite{Dehnen2001} and a technique to subtract the uniform background density, the error behavior and accuracy are particularly adapted for cosmological volumes (see \cite{Sutter2014f} for details). To generate initial conditions we used a power spectrum calculated with CLASS \cite{Blas2011}, and then realized the initial conditions with a modified version of 2LPTIC \cite{Crocce2006}. The simulations assume the fiducial Planck cosmology, we have then independently varied $w_0$ and $w_a$ while adapting the amplitude of initial fluctuations such as to preserve the value of $\sigma_8$ determined by Planck. The size of the box is 1 $Gpch^{-1}$ and it contains $1024^3$ particles. We subsample the simulation to a mean density of $\bar{n}=4 \times 10^{-3}$ particles per cubic $h^{-1}Mpc$, comparable to the number density expected from future surveys \cite{Sutter2014f}. 

The comparison of the void abundance in the \lcdm simulation with the theoretical model is shown in Figure \ref{fig: Abundances0}. The theoretical model is able to match the abundances from simulations, except for smaller radius which we do not use to constrain parameters (as discussed in Section III).
To parametrize the uncertainty in the determination of the parameter $\delta _v$ we can marginalize over it to obtain the constraints. The shaded areas in Figure \ref{fig: Abundances0} show the effects of the variation of the value of $\delta_{v}$ (varied by 0.05 and 0.1 in the Figure) for the \lcdm case. 
The marginalization over $\delta_{v}$ allows to take into account the fact that the Sheth and Van de Weygaert fit is qualitative, and only roughly matches simulations, due to the discreteness of the galaxy distribution as opposed to a smooth density field.

Figures \ref{fig: Abundances1} and \ref{fig: Abundances2} show the comparison of void abundance with simulations for different values of the parameters $w_{0}$ and $w_{a}$. The used simulations mimic the number density of tracers in the two surveys, thus allowing us to match the abundance of voids in a realistic framework (although the volumes of the simulations are smaller than the actual volume to be obtained with both surveys). While at larger radii we are able to match the behaviour of simulations, we see that in Figures \ref{fig: Abundances1} and \ref{fig: Abundances2} the theoretical model and the curves from simulations diverge at small radii. 
For voids of small radii, the void number function increases dramatically. As described in Section III, to ensure a robust estimation for the number of voids, we cut off the constraints only considering voids at higher radii and thus avoiding to consider voids at smaller radii.

Additionally, as mentioned before (and shown in Figure \ref{fig: Abundances0}), in order to account for the residual uncertainty in the Sheth and Van de Weygaert fit, we will marginalize over the parameter $\delta_{v}$. 
The abundance of voids obtained with the described methodology gives a measure of the power of the survey to constrain cosmological parameters, and is particularly robust, as based on recent observations.
The next Section presents the results of the calculation.

\section{Results}

\begin{figure}
\centering
\epsfig{file=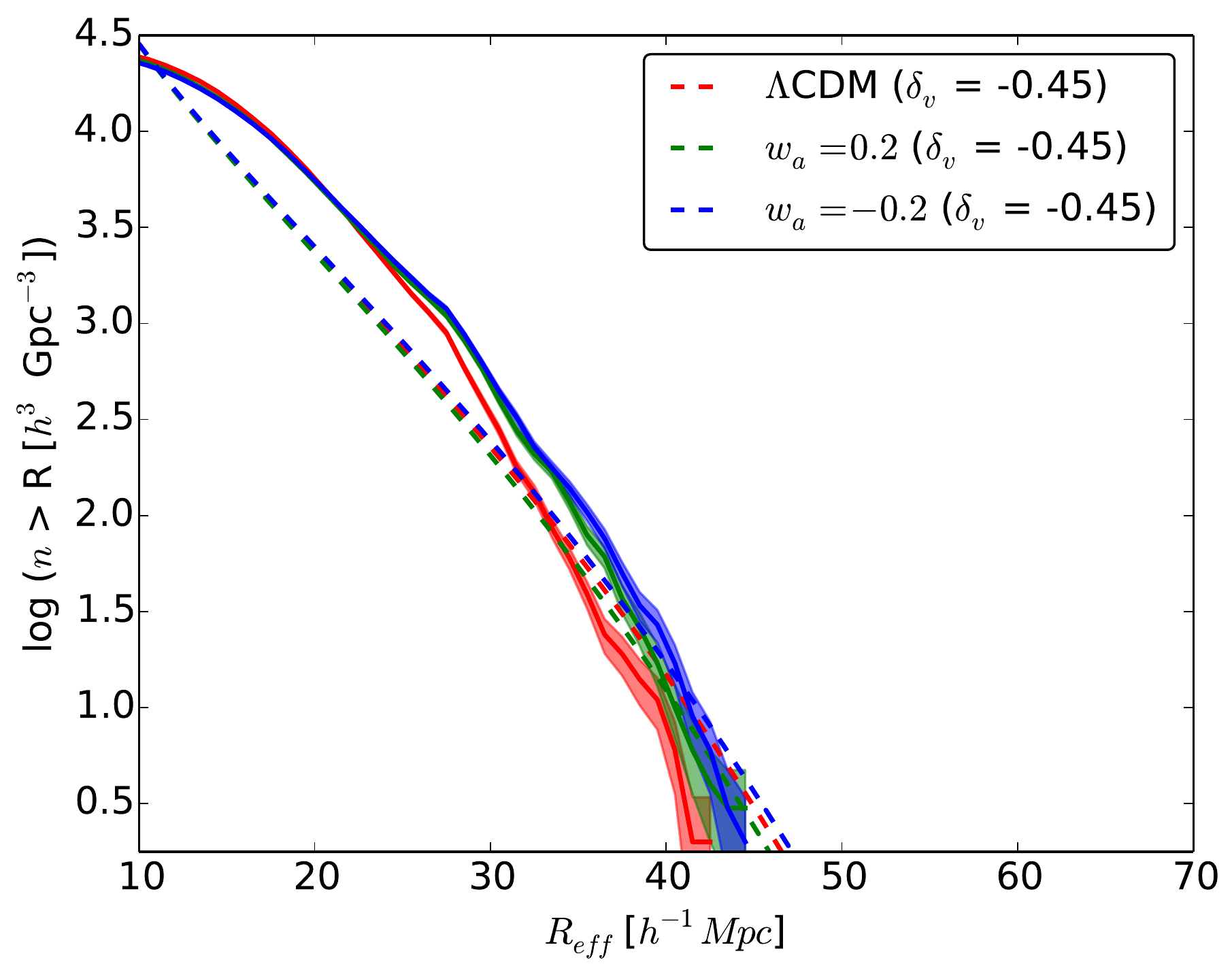,width=1\linewidth,clip=, angle=0}  
\caption{Matching of the abundance in simulations for models with different $w_a$. The thick lines show the abundances from simulations for models with different $w_a$ (and fixed $w_0$). The prediction from the Sheth and Van de Weygaert model are represented with dashed lines. 
 \label{fig: Abundances2}}
\end{figure}

The cosmological parameters we use to evaluate the Fisher matrix together with their fiducial values are given in table \ref{PriorsTable} (and correspond to Planck values \cite{PlanckCollaboration2014}). We use the BBKS analytical fit \cite{Bardeen1986} for the transfer function $T(k)$, and consider its generalization \cite{Eisenstein1999} that takes the effect of baryons into account. The matter power spectrum at any given redshift can then be calculated as

\begin{equation}
P_{\delta}(k,a)=\frac{4}{25} T^2(k) D^2(a) \frac{k^4 \mathrm{c}^4}{\Omega_{m0}^2 H_0^4} P_{\zeta}(k)
\end{equation}

\noindent Here, $P_{\zeta}$ is the power spectrum of the primordial curvature perturbation and is assumed to be of the form $\Delta_{\zeta}^2(k) \equiv \frac{k^3 P_{\zeta}(k)}{2 \pi^2} = \Delta_{\zeta}^2(k_0) (k/k_0)^{n_\mathrm{s}(k_0)-1}$, where $k_0~=~0.05~h\mathrm{Mpc}^{-1}$ is the pivot scale. 
Also, $D(a)$ is the linear growth factor of structure, normalized so that at late time $D(a=1)=1$. In the case of a non-clustering dark energy fluid it can be found as the solution of the following differential equation:
\begin{eqnarray} \label{eq: diff}
&& \frac{3}{2} \Omega_m(a) = \frac{\mathrm{d}^2 \ln \delta}{\mathrm{d} (\ln a)^2} + \left( \frac{\mathrm{d}\ln \delta}{\mathrm{d} \ln a} \right)^2 +\\
\nonumber && + \frac{\mathrm{d} \ln \delta}{\mathrm{d} \ln a} \left( 1-\frac{1}{2} \left[ \Omega_\mathrm{m}(a)+(3 w(a)+1) \Omega_\mathrm{DE}(a) \right] \right);
\end{eqnarray}
from \cite{Percival2005} (to be more precise, $D(a)$ is the growing mode of the solution of equation \ref{eq: diff}, we have $\delta \propto D(a) $).

\begin{table}
\caption{Parametrization of the cosmology and the fiducial values chosen
for the maximal set of parameters used in evaluating the Fisher forecasts.}
\begin{center}
\begin{tabular}{|c|c|c|c|c|c|c|c|c|}
\hline
$\Omega_{\rm{b}}$ & $\Omega_{\rm{m}}$
& $h$ & $\tau$ & $\Omega_{\rm{k}}$ & $w_0$ &$w_a$ & $n_s$ &
 $ln(10^{10}\Delta_{\zeta}(k_0))$ \\
\hline
 0.049 & 0.318 & 0.67& 0.09 & 0.0 & -1 & 0 & 0.96 & 3.098 \\
\hline
\end{tabular}
\end{center}
\label{PriorsTable}
\end{table}

To find out how a combination of different surveys can constrain the cosmological parameters, we simply need to add the Fisher matrices from those surveys. 
However, it is important to have the same set of cosmological parameters for all surveys. We can change from one set of parameters to others by using $\mathbf{F}'=\mathbf{J}^{\dagger}.\mathbf{F}.\mathbf{J}$ in which $J_{ij} \equiv \partial{\theta_i}/\partial{\theta'_j}$ is the Jacobian of the transformation.

We consider Fisher forecasts computed from the Planck satellite (note that this forecast does not include CMB lensing); SN is the Fisher matrix for a Type Ia supernova survey such as the Large Synoptic Survey Telescope \cite{LSST2009,Biswas2010}. 

Inverting the Fisher matrices, we can obtain the 1-$\sigma$ error ellipses in the $w_0 - w_a$ plane, that measure the capacity of a combination of experiments to constrain the dark energy equation of state. 

\begin{figure}
\centering
\epsfig{file=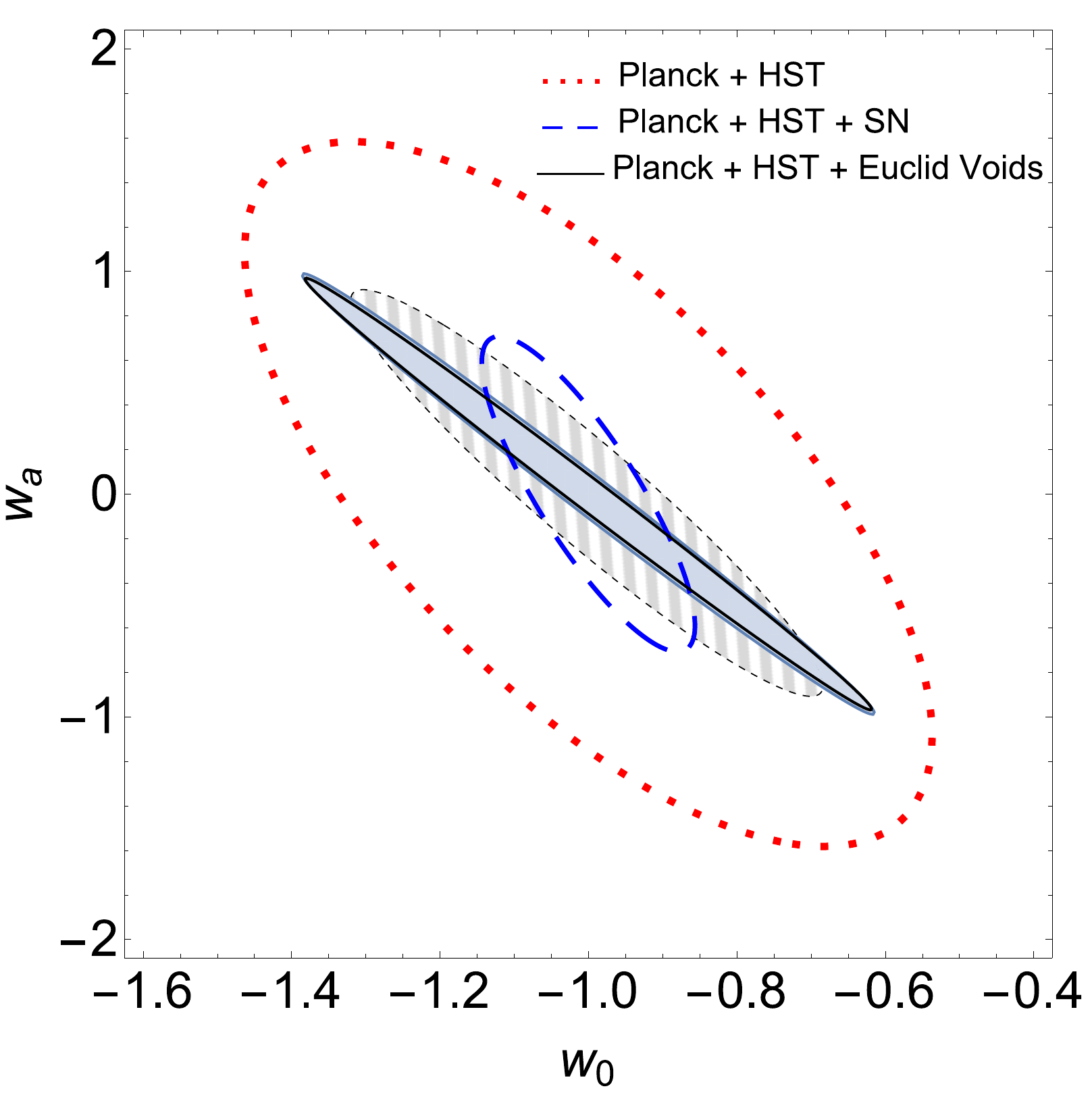,width=0.93\linewidth,clip=, angle=0}  
\caption{1-$\sigma$ error ellipses for several combinations of data sets. From outer ellipse to inner: Planck+HST, Planck+HST+SN, Planck+HST+Euclid Voids. The light-blue filled contour shows the result of the marginalization over the parameter $\delta_{v}$. For comparison, we also show in the striped-filled contour the BAO constraints from Euclid (also added to Planck+HST), from \cite{Lavaux2012}.\label{fig: Fisher Matrix EU}}
\end{figure}

\begin{figure}
\centering
\epsfig{file=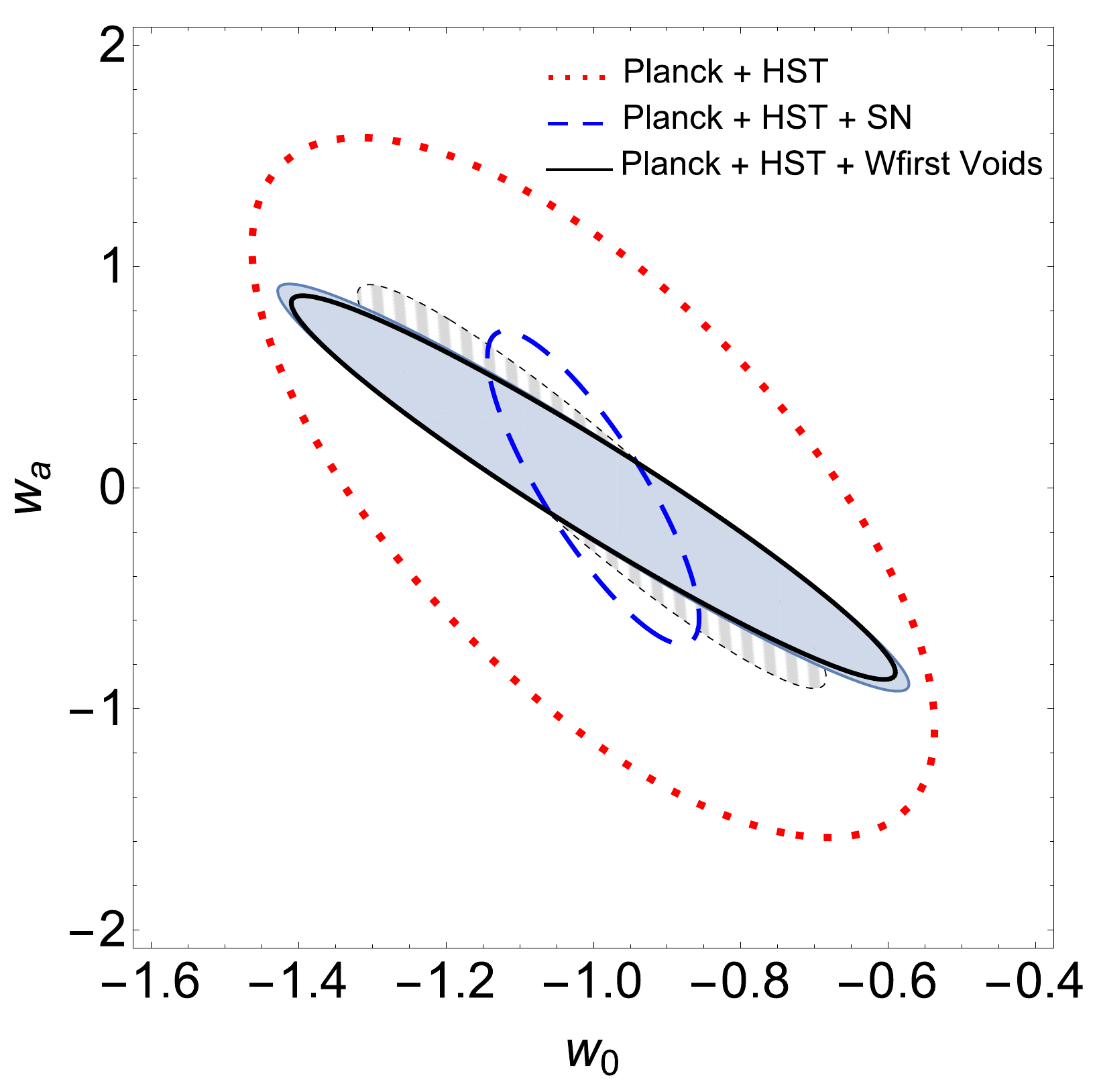,width=0.93\linewidth,clip=, angle=0}  
\caption{1-$\sigma$ error ellipses for several combination of data sets. From outer ellipse to inner: Planck+HST, Planck+HST+SN, Planck+HST+WFIRST Voids. The light-blue filled contour shows the result of the marginalization over the parameter $\delta_{v}$. For comparison, we also show in the striped-filled contour the BAO constraints from Euclid (also added to Planck+HST), from \cite{Lavaux2012}.\label{fig: Fisher Matrix WFIRST}}
\end{figure}

We show, respectively in Figures \ref{fig: Fisher Matrix EU} and \ref{fig: Fisher Matrix WFIRST}, the 1-$\sigma$ error ellipses in the $w_0 - w_a$ plane for Euclid and WFIRST voids combined with Planck and HST (namely the Hubble Space Telescope Survey measurements of the Hubble constant from \cite{Freedman2001}). For comparison, we also show the ellipse of Planck+HST alone, the Planck+HST+ SN and the Planck + HST+ Euclid BAO (from \cite{Lavaux2012}). Figures \ref{fig: Fisher Matrix EU} and \ref{fig: Fisher Matrix WFIRST} thus allow us to compare the constraining power of void abundances for Euclid and WFIRST with BAO and supernovae. 

Additionally, to parametrize the uncertainty on the $\delta_{v}$ parameter, we can marginalize over it (as well as marginalizing over standard parameters). We vary the parameter $\delta_{v}$ by $\pm0.05$, thus taking into account the effects of sparsity (due to the discreteness of the galaxy distribution as opposed to a smooth density field) that might affect the void abundance and the ordering discrepancies between the simulations and the Sheth and Van de Weygaert fit seen in Figure \ref{fig: Abundances2}. 
For the marginalisation to characterize the uncertainties and biases between the Sheth and Van de Weygaert model and the simulations we use a uniform prior ($-0.45 \pm 0.05$). While we don't expect our results to change substantially in the case of e.g. a gaussian prior (because the cosmological constraints are relatively robust with respect to the uncertainty in $\delta_v$), the precision in the calculation of the expected constraints is currently more strongly affected by the uncertainty about the details of the surveys than by the choice of the prior. Even though it is not the purpose of this paper to perform a full Markov Chain Monte Carlo exploration of the probability distribution, we point out that it would be the optimal way to explore the probability distribution function to be considered for future work, when details of the surveys will be updated and thus more reliable, allowing a sensitive improvement in the precision of constraints.
The marginalization over $\delta_{v}$ also allows to take into account the residual effects of peculiar velocities on void abundance for voids of larger size, as well as further uncertainties related to the watershed definition of voids in the tracer distribution \cite{Furlanetto2006}. 
Figures \ref{fig: Fisher Matrix EU} and \ref{fig: Fisher Matrix WFIRST} also show (light-blue filled region) the 1-$\sigma$ error ellipses in the $w_0 - w_a$ plane for the satellites after marginalization. We observe that the ellipse is only slightly thickened, showing that the cosmological constraints are relatively robust with respect to the uncertainty in $\delta_{v}$.

\begin{figure}[h]
\centering
\epsfig{file=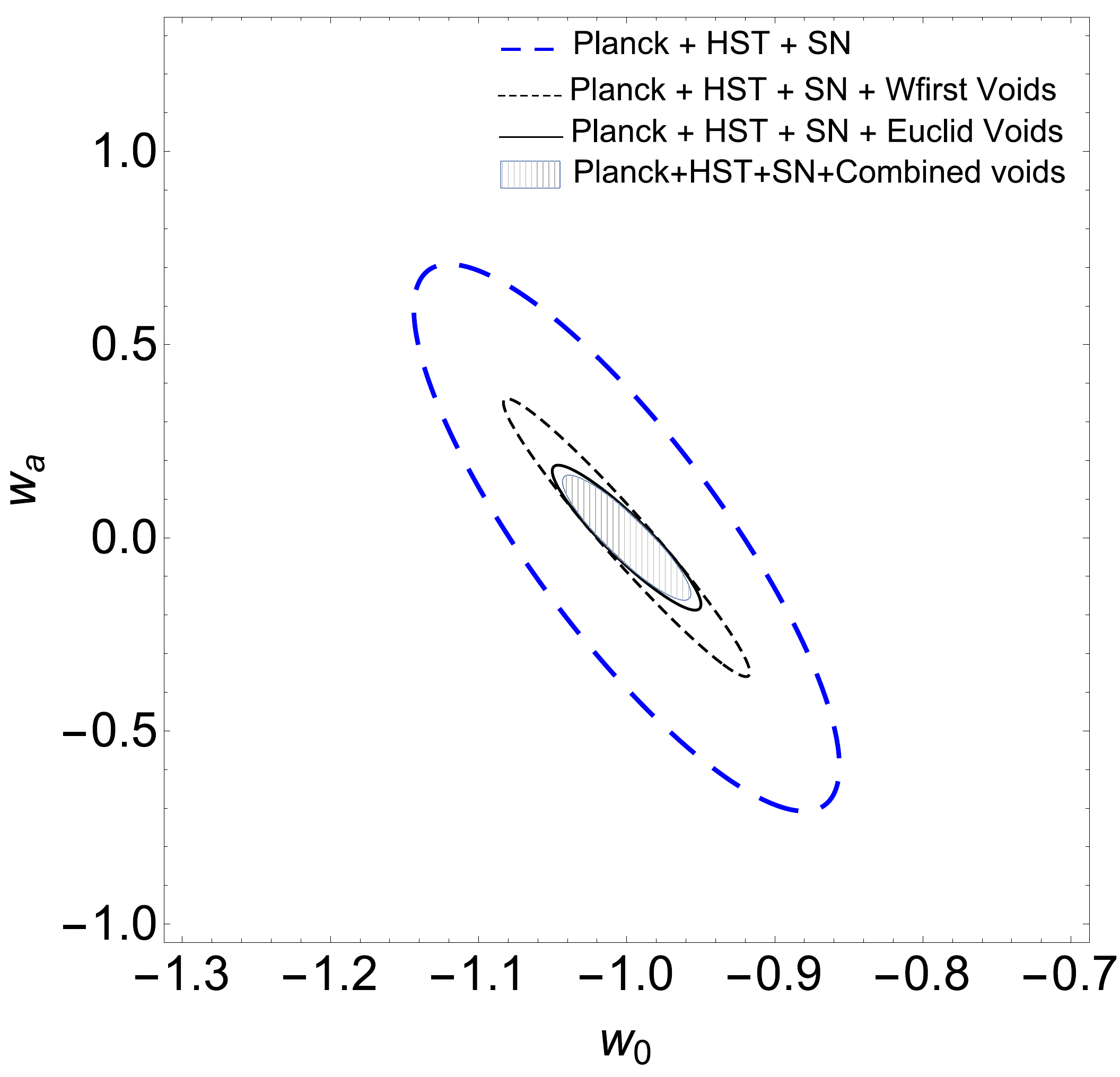,width=1\linewidth,clip=, angle=0}  
\caption{Comparison of 1-$\sigma$ error ellipses for WFIRST and Euclid voids when the supernovae data is also considered (in each case the void abundance constraints are added to Planck+HST+SN data). We also show the constraints obtained by considering an estimate of the combination of voids from both surveys (see Section IV). (It should be noted that the axis range is smaller in this Figure.) \label{fig: Comparison}}
\end{figure}

Interestingly, we notice that the void abundance with WFIRST, while leading to similar constraints on the parameter $w_{0}$, constrains the $w_{a}$ parameter slightly better than Euclid. Nevertheless, the conditional error on $w_{a}$ from Euclid is lower -- this means that if we combine constraints from void abundances with other probes that constrain effectively $w_{0}$ (such as supernovae), we can expect the resulting combined constraints on $w_{a}$ to be better for the probe combined with Euclid (as we will see in Figure  \ref{fig: Comparison}). 

Indeed, so far we have only compared the constraints from the void abundance to the contraints from supernovae, but additional constraints are obtained if supernovae are considered jointly with voids. Figure \ref{fig: Comparison} shows the comparison of WFIRST and Euclid void constraints when both are added to Planck + HST + SN. It should be noted that the axis range is smaller in this Figure.

When independently added to supernovae, the void abundance from Euclid and WFIRST results in an enhanced constraint, although, as expected from the previous discussion, Euclid combined with supernovae is most constraining.
Both WFIRST and LSST promise to bring new SN data. We can expect that, when added to the complementary probe from void abundance, these data increase the quality of constraints. While voids alone already bring important information to constrain the dark energy equation of state, the different orientation of ellipses -- related to the different nature of probes as well as to the different systematics affecting the probes -- for SN and voids will be a consistent asset to fully exploit the complementarity of the two probes.
The resulting allowed $w_{0}-w_{a}$ parameter space is considerably reduced when voids are added to the supernovae probe, thus showing the power of void number counts to constrain the dark energy equation of state.  

\begin{table}[h]
\caption{Total number of voids forecasted for each survey.}
\begin{center}
\begin{tabular}{|c|c|c|}
\hline
Survey& Sky Fraction &Total Number of voids  \\
\hline
Euclid & 0.36 & 7.8$\times 10^{5}$\\
\hline
WFIRST &0.05 & 2.5$ \times 10^{5}$\\
\hline
\end{tabular}
\end{center}
\label{tab: Total Voids}
\end{table}

Also, it is worth notice that, while WFIRST will find less voids than Euclid (see Table \ref{tab: Total Voids}), it still brings a competitive constraint as far as the void abundance is concerned, because it is sampling the cosmic web at smaller scales (higher density). 

The error ellipses of the two surveys cover different parts of the parameter space; thus --- although it remains non trivial to combine cosmological constraints from the two data sets \cite{Jain2015} --- if jointly considered, they promise to yield improved constraints on the $w_{0}-w_{a}$ plane. The different constraints on the parameter space from the two surveys can be understood considering that they have galaxy number densities peaked at different redshifts (namely, the Euclid peak galaxy number density will be at $z\simeq 0.75-0.85$, while the WFIRST peak will be at $z\simeq 1.30-1.45$).

In addition to having different degeneracy directions, the Euclid and WFIRST surveys will also have different observational systematic errors due to their different observing strategies. In general, a space-based slitless redshift survey possesses depth variations across the field of view of the instrument due to varying image quality, spectral resolution, and coverage (chip gaps and defects), and over the area of the survey due to variations in sky brightness and stellar confusion. 

The WFIRST tiling strategy results in most galaxies being observed multiple times at widely separated positions on the focal plane and at different zodiacal brightness levels, which will help to mitigate effects that depend on field position and sky brightness. However, its warm telescope results in a position-dependent thermal background that is highest near the center of the focal plane. 

Both surveys use multiple dispersion angles (implemented on Euclid by changing the grism and WFIRST by rotating the entire telescope). While we leave an investigation of these effects on void statistics to future work, we note that comparison of the data sets in their region of overlap should help to constrain the systematics models for both surveys.

As mentioned, because of correlations it would be non-trivial to combine Euclid and WFIRST constraints in the range where the two surveys overlap; nevertheless we can obtain a close estimate of the combined power of void abundances from the two surveys by considering that WFIRST is deeper that Euclid in the covered space, and Euclid is much wider. Thus as a simple estimation we can consider constraints from a survey as Euclid but with a smaller volume (i.e. subtracting the volume of WFIRST) and then adding the constraints of this survey to the constraints from WFIRST. We show the estimate of constraints obtained with such methodology in Figure \ref{fig: Comparison}. As expected, this estimate shows that combining voids from the two surveys would allow to enhance the quality of constraints.

We showed that the large volume and deepness of the surveys allows to constrain effectively the $w_{0}-w_{a}$ parameter space by using a phenomenological prediction for the number of voids with Euclid and WFIRST based on simulations. To further dramatically improve the potential of the method, one could also consider voids at smaller radii which have not been used in our forecast -- this would need a theoretical model able to predict the abundance more in detail also for smaller voids, and the use of simulations with larger volumes to confirm predictions \cite{Chongchitnan2015}. Since the abundance of voids increases rapidly with decreasing radius, this could lead to a powerful increase in statistics and thus in even better constraints.
The results we obtain so far are able to show the potential of the method and serve as an initial probe of the power of these surveys in the framework of void abundance.

\section{Conclusion}
In this paper we have considered the information gained by using the number of observed voids as a function of redshift to constrain the equation of state of dark energy. The void number count is well performing in terms of improving constraints on $w(z)$, in particular when combined with other probes. The advantage of void number counts is that the method does not need a separate expensive survey, since it can use data from the existing or planned redshift surveys to give extra information about the nature of dark energy.

The method of observed cluster counts has been used to constrain the equation of state of dark energy (or, more generally, cosmological models, \textit{e.g.} \cite{Wang2004,Abramo2009,Basilakos2010,Campanelli2012,Devi2014}). It is worth noticing that the observed void counts promise to give complementary constraints to the cluster method. The first advantage of considering the void method is that the systematics will be different than in the cluster case. The estimation of the number of voids is based on the void's observed ``geometry" (radius), while the method for clusters relies on the use of the cluster ``masses". 

Clearly, there are different systematic errors involved in their estimation, and the confidence of the detection of a signal will be greatly enhanced when they are combined. Additionally, as mentioned, the measurement of voids does not need a different survey. To extract cosmological information it is possible to directly use surveys already mapping the cosmic web for probes such as BAO --- thus obtaining a complementary constraint. 

The second advantage is that the linear density threshold for void formation is different than the one for halos, which means we are considering a different probe. Finally, the redshift dependences of the number of clusters and voids are different, since the mass scales probed are different. The abundance of clusters and voids are thus sensitive to the equation of state of dark energy in different redshift ranges. For such reason the use of voids can give complementary and independent constraints on dark energy.

We have compared two upcoming satellites: Euclid and WFIRST; and we have showed that their measured void abundance will constrain the parameter space $w_{0}-w_{a}$ differently, thus optimally sampling the cosmic web at a different scale and redshift. This results in joint constraints of an increased precision. Finally, the allowed $w_{0}-w_{a}$ parameter space is further reduced when void constraints are added to the supernovae probe, thus showing the power of the void number counts method to constrain the dark energy equation of state. \\

\vspace{10pt}
\section*{Acknowledgments}
\vspace{10pt}

AP thanks St\'ephanie Escoffier, Emanuele Castorina and Andr\'e Tilquin for useful discussions. AP acknowledges financial support from the grant OMEGA ANR-11-JS56- 003-01. AP and BDW acknowledge support from BDW's Chaire Internationale in Theoretical Cosmology at the Universit\'{e} Pierre et Marie Curie. PMS is supported by the INFN IS PD51 ``Indark". PMS and BDW acknowledge support from NSF Grant AST-0908902. RB acknowledges partial support from the Washington Research Foundation Fund for Innovation in Data-Intensive Discovery and the Moore/Sloan Data Science Environments Project at the University of Washington. BDW is partially supported by a senior Excellence Chair of the Agence Nationale de la Recherche (ANR-10-CEXC-004- 01). CMH is supported by the US Department of Energy under contract DE-FG03-02-ER40701, the David and Lucile Packard Foundation, and the Simons Foundation. 
We thank the referee for the comments that allowed us to clarify some important points in this work. This work made use of the Horizon Cluster at the Institut d'Astrophysique de Paris; AP particularly thanks Christophe Pichon for support. This research is partially supported by NSF AST 09-08693 ARRA.

\vspace{40pt}

\bibliographystyle{h-physrev}
\bibliography{paper}

\end{document}